\documentstyle[aps,preprint,epsf]{revtex}
\begin{document}
\draft
%\twocolumn[\hsize\textwidth\columnwidth\hsize\csname@twocolumnfalse\endcsname
%\maketitle
\title{Antiferromagnetism and phase separation
in electronic models for doped transition-metal oxides}
\author{Shun-Qing Shen and Z. D. Wang}
\address{Department of Physics, The University of Hong Kong,
Pokfulam Road, Hong Kong}
%\date{March 12, 1998 }
\maketitle
%\twocolumn[\hsize\textwidth\columnwidth\hsize\csname@twocolumnfalse\endcsname
\begin{abstract}
We investigate the ground state properties of electronic models for doped
transition-metal oxides. An effective t -  J like Hamiltonian is
derived from the case of strong Hund coupling between the conduction
electrons and localized spins by means of the projection technique.
An attractive interaction for conduction electrons and
an antiferromagnetic coupling of the localized spin are obtained.
A large ratio of the attraction to
effective electron hopping, which is modulated by the spin background,
will lead to the phase separation. The antiferromagnetic phase and the
phase separation appear in the case of either high or low density of
electrons. The possible relevance of the phase separation to the
charge stripe phase in doped transition-metal oxides is discussed. 
\end{abstract}
                                                 
%]

The problem of doped Mott insulators has attracted much attention
because of its
relevance to high temperature superconductivity and colossal magnetoresistance
effect. Recent experiments of doped lanthanum cuperate \cite{Tranquada95},
nickelate \cite{Hayden92} and
manganite \cite{Jirak85}
families of materials exhibit a new type of charge ordering and
spin ordering in an extensive region.
For example, the charge and spin stripe phases were observed in
La$_{2-x}$Sr$_x$NiO$_4$ samples \cite{Blumberg98}. Along the charge stripe,
there is
strong antiferromagnetic correlation.
It is also shown experimentally
that the charge ordering collapses in the presence of an external
magnetic field, which can destroy antiferromagnetic ordering\cite{Tomioka95}.
Many efforts have been devoted to understand the origin of the phenomena and its intrinsic relevance to
various anomalous transport properties.

In this paper, starting from an electronic model for doped transition-metal oxides, we derive an effective t-J like Hamiltonian
for the case of
strong Hund coupling. An attractive interaction between conduction
electrons, which is associated with the antiferromagnetic correlation,
is obtained. Both the attraction and electron hopping are modulated
by the configuration of two localized spins on the nearest neighbor sites.
A larger ratio between them will lead to the phase separation, which
is expected in terms of the ideas of frustrated phase separation
for the charge stripe phase \cite{Low94}. An antiferromagnetic background
leads to
attraction between electrons. We find that the phase separation with
electron-rich and -poor regimes has a lower energy than an antiferromagnetic
phase with a uniform density of charge. A phase diagram for the model is presented.
The possible relevance to the phase separation and the charge stripe phases in doped
lanthanum manganites and nickelates are also discussed.

An electronic model to describe doped transition metal oxides is a Kondo-like
lattice Hamiltonian with the strong Hund coupling
\begin{equation}
H = -t \sum_{\langle ij \rangle, \sigma} c^{\dagger}_{i,\sigma} c_{j,\sigma}
- J_H \sum_{i} {\bf S}_{i}\cdot{\bf S}_{ic}
\label{kondo-model}
\end{equation}
where $c^{\dagger}_{i,\sigma}$ and $c_{i,\sigma}$ are the creation and
annihilation operators for conduction electrons, respectively.
${\bf S}_{ic} = \sum_{\sigma,\sigma'}{\bf \sigma}_{\sigma,\sigma'}
c^{\dagger}_{i,\sigma}c_{i,\sigma'}/2$ is the spin operator for the
conduction electron and ${\bf \sigma}$ are the Pauli matrices.
${\bf S}_i$ is the total spin of the localized electrons at site $i$.
$J_H >0$ is the Hund coupling between the conduction and localized
electrons. In the manganites, three $t_{2g}$ electrons are almost localized
and form an $S=3/2$ spin state according to the Hund rule. Electrons in 
$e_g$ orbital form a conduction band \cite{Coey95}.  In the nickelate, the
localized spin is just $S=1/2$ \cite{Kuiper89}. In the case of single electron the ground state
is a fully saturated ferromagnet. It is easier for the conduction
electron to move when the two
localized spins on the nearest neighbor sites are parallel to each other.
The process may lead to metallic ferromagnetism, and is called the  double
exchange mechanism \cite{Zener51,Anderson55,Gennes60}.

Usually the Hund coupling is very large in either manganites or nickelates.
An infinite $J_H$ limit is often taken in these systems, especially to investigate
the double exchange ferromagnetic phase. However, in the limit, the spin of electron
is completely frozen to localized spin to form an $S+1/2$ state due to the
strong Hund coupling, and the model is reduced to a spinless
fermion system without a direct
electron-electron interaction, which cannot describe
the charge ordering and antiferromagnetism. We consider the large and finite
$J_H$ ($ \gg t$) case. As the strong Hund coupling forces most of electrons to form
$S+1/2$ states with the localized spins, we will restrict our discussion in the
space, which includes only the empty and single occupancies with $S+1/2$ state.
The finite $J_H$ effect  can be regarded as the perturbation correction to the
large $J_H$ limit.  The operator to project onto the space of the
configurations with
empty and the $S+1/2$ states is
\begin{equation}
P = \prod_i P_i = \prod_i (P_{hi} + P_{si}^+)
\end{equation} 
where $P_{hi} = (1-n_{i,\uparrow})(1-n_{i,\downarrow})$ and 
\begin{eqnarray}
P_{si}^+ &=& \sum_{\sigma,\sigma'} (P_{i}^+)_{\sigma\sigma'}
\tilde{c}^{\dagger}_{i,\sigma}\tilde{c}_{i,\sigma'}\nonumber \\
&=& \sum_{\sigma,\sigma'} 
\left (
\frac{{\bf S}_i \cdot {\bf \sigma} + (S +1) {\bf I}}{2S+1}
\right )_{\sigma\sigma'} \tilde{c}^{\dagger}_{i,\sigma}
\tilde{c}_{i,\sigma'}. \nonumber 
\end{eqnarray}
The operator
$\tilde{c}^{\dagger}_{i,\sigma} =
(1-n_{i,-\sigma})c^{\dagger}_{i,\sigma}$
rules out double
occupancy on the same site. ${\bf I}$ is a unity matrix. 
Utilizing the Schrieffer-Wolf transformation \cite{Schrieffer66},
a t-J like effective Hamiltonian is derived
\begin{equation}
H_{eff} \approx -t\sum_{\langle ij \rangle} 
\bar{c}^{\dagger}_{i,\sigma}\bar{c}_{j,\sigma}+
4v_0\sum_{\langle ij\rangle}
( \bar{\bf S}_{ic}\cdot\bar{\bf S}_{jc} - \frac{1}{4}\bar{n}_{i}\bar{n}_j)
\label{tj-model},
\end{equation}
where $v_0 = t^2/(J_HS)$,
$\bar{\bf S}_{ic}  = \sum_{\sigma,\sigma'}
({\bf \sigma})_{\sigma\sigma'}
\bar{c}^{\dagger}_{i,\sigma}\bar{c}_{i,\sigma'}/2$, and
$\bar{c}_{i,\sigma} =
\sum_{\sigma'} (P_i^+)_{\sigma\sigma'}\tilde{c}_{i,\sigma'}$.
In Eq.(\ref{tj-model}), except for ignorance of
higher order perturbation correction and
a constant term $-N_e J_H S/2$ ($N_e$ is the number of electrons), we also
neglect three-site terms in order of $t^2/J_HS$ which describe indirect
hopping process between the next nearest neighbor sites. They are of order
$t/J_HS$ ($\ll 1$) when compared with the first, direct hopping term in Eq. (3).
A detailed discussion including these terms will be published elsewhere.
The physical meaning of the operator $\bar{\bf S}_{ic}$ is the component
of electron spin along the localized spin on the same site meanwhile the
electron and localized spin form an spin $S+1/2$ state. It is shown that
$(2S+1)\bar{\bf S}_{ic}$ (=${\bf S}^t_{i}$)
is an spin operator with $S+1/2$ if the site is
occupied by a single electron.
Let us first consider two limits. When $J_H\rightarrow +\infty$,
$v_0\rightarrow 0$.
The model is reduced to the quantum double exchange model
\cite{Kubo72,Shen97}. Expanding the dressed operators $\bar{c}$ in
Eq.(\ref{tj-model}), we find
a direct exchange term for localized spins and its effective exchange
coupling is approximately 
$$J_{de} \approx -t \sum_{\sigma} \langle
\tilde{c}_{i,\sigma}^{\dagger} \tilde{c}_{j,\sigma}\rangle/(2S+1)^2
$$
where $\langle \dots\rangle$ represents the average of the ground state.
The coupling is proportional to the kinetic energy and is always
ferromagnetic. It reaches its minimum at quarter filling ($\rho =1/2$)
and vanishes at two density limits
$\rho =0$ and 1 in a sense of the mean field approximation. This result
consists with the physical picture of the
double exchange ferromagnet.
At half filling, which means that the number of electrons is equal
to the number
of lattice sites, each site is occupied by one electron and
there is no empty. The effective Hamiltonian (\ref{tj-model}) is reduced to
an antiferromagnetic Heisenberg model with spin $S+1/2$
\begin{equation}
H_{AF}= \frac{4v_0}{(2S+1)^2}  \sum_{\langle ij\rangle}
\left ( {\bf S}_{i}^t\cdot{\bf S}_{j}^t - (S+\frac{1}{2})^2 \right).
\label{heisenberg}
\end{equation}
This is consistent with rigorous results of the model (\ref{kondo-model}) at
half filling for any $J_H$ that the ground state is spin singlet on a hyper-cubic
lattice \cite{Shen96}. In the antiferromagnetic Heisenberg model,
it is shown rigorously that
the ground state possesses antiferromagnetic long-range order on a square
lattice for spin 1 or higher and cubic
lattice for spin $1/2$ or higher \cite{Dyson78}. Therefore, here as
$S+1/2 >1$, we conclude that the Kondo lattice model
at half filling and with strong Hund coupling possesses antiferromagnetic long-range
order,  which is completely opposite to the case of low density of electrons
where ferromagnetic correlation is predominant.

Although the physics of the two terms in the Hamiltonian (\ref{tj-model}) is
clear, the combination of the two terms make it very complicated.
The usual t-J model from the large U Hubbard model can be regarded as a specific
case of $S=0$ with a finite $v_0$. Many efforts are
attempted to investigate the antiferromagnetism and superconductivity.
Emery {\it et  al.}
proposed that a Heisenberg antiferromagnet in t-J model is always unstable to a phase
separation at a sufficient dilute doping \cite{Emery90}.
Their conjecture is supported numerically at larger $J/t$, but
it is still open problem at small $J/t$ strength
\cite{Hellberg97}.
Nevertheless, for the usual t-J model, $J=4t^2/U$ and
should be very small in a physical region. The localized spin in J-term
is 1/2. In our case, the spin background can modulate the
electronic behavior, and localized spin can be any value. To simplify our discussion, we take the classic spin
approximation or large S limit. 
The spin ${\bf S}_i$
can be parameterized by polar angles $\theta_i$ and $\phi_i$ and
$S/(2S+1)\approx 1/2$. ${\bf S}_i = S \vec{s}_{0i}$ and
$\vec{s}_{0i} = (\sin\theta_i\cos\phi_i,
\sin\theta_i\sin\phi_i,\cos\theta_i)$.
Except for the exchange coupling between the conduction electron and
localized spin, it is
believed that the antiferromagnetic exchange coupling between the
localized spins
$J_{AF} {\bf S}_i \cdot {\bf S}_j$ also plays an essential role in
determining the phase
diagram of manganites, especially in the region of low density
of electrons. The projected operator
$\bar{\bf S}_i = P_i {\bf S}_i P_i = S\vec{s}_{0i}.$
We shall consider it in our following discussion.
In this approach, the quantum t-J model including the
localized spin coupling
is reduced to 
\begin{eqnarray}
H_{rde} = &-& t \sum_{\langle ij \rangle} c_{ij}
\alpha^{\dagger}_{i} \alpha_{j} - 2v_0\sum_{\langle ij \rangle}
\sin^2\frac{\Theta_{ij}}{2} \alpha^{\dagger}_{i}
\alpha_{i} \alpha^{\dagger}_{j} \alpha_{j} \nonumber \\
&+& J_{AF}S^2\sum_{\langle ij \rangle} \cos\Theta_{ij}
\label{rde-model}
\end{eqnarray}
where 
$\alpha_i = \cos\frac{\theta_i}{2} \tilde{c}_{i,\uparrow} + \sin\frac{\theta_i}{2}e^{-i\phi_i}
\tilde{c}_{i,\downarrow}$;
$$c_{ij}
= \cos \frac{\theta_i}{2}\cos\frac{\theta_j}{2}
+ \sin\frac{\theta_i}{2}\sin\frac{\theta_j}{2} e^{-i(\phi_i -\phi_j)};$$
$$\cos\Theta_{ij} = \cos \theta_i \cos\theta_j
+ \sin\theta_i\sin\theta_j \cos(\phi_i -\phi_j).$$
$\Theta_{ij}$ is the angle between the two spin units $\vec{s}_{0i}$ and $\vec{s}_{0j}$ and
$\vert c_{ij} \vert = \cos (\Theta_{ij}/2)$.
$\alpha$ operators are for the conduction electrons whose spins are frozen by the localized 
spins on the same site, and therefore can be considered only
to describe the charge degrees of freedom.
The first part of $H_{rde}$ is the usual
double exchange model with a Berry phase,
and the second part comes from the correction of the finite $J_H$.
Both the renormalized coefficients $c_{ij}$ of hopping terms and
$\sin^2(\Theta_{ij}/2)$ of the
density-density interaction depend on the background of the spin
configurations.   Our following discussion will be based on the Hamiltonian
(\ref{rde-model}).

From the point of view of localized spins,
the mobile electrons favor to the ferromagnetic correlation. However, the
finite $J_H$ as well as the direct exchange coupling $J_{AF}$ tends to form
antiferromagnetism. For instance, in the one dimensional case, the effective
double exchange coupling is approximately proportional to $-\sin\rho\pi/\pi$.
It reaches
its minimum at quarter filling and approaches to zero at two end limits. The
exchange coupling from the finite $J_H$ is approximately
proportional to $\rho^2$.
It is stronger than
the double exchange
coupling at a higher density of electrons, but is weaker at a lower
density of electrons. As far as the direct exchange coupling $J_{AF}S^2$
is introduced, the double exchange coupling is always suppressed
at the two end limits of density. If $J_{AF}S^2$ is sufficiently large, the
ferromagnetic phase is always unstable.
If $J_{AF}S^2$  is not sufficiently large the double exchange ferromagnetism can survive
in a finite range of doping.
A possible phase diagram is shown in Fig. 1 for the
chosen parameters. As
$\vert c_{ij}\vert$ is proportional to $\cos(\Theta/2)$, not $\cos\Theta$ for
the ferromagnetic coupling, it is also possible to lead some non-collinear
magnetism \cite{Gennes60,Shen97,Inoue95}. The boundaries of phases in
Fig.1 and later in Figs. 2 and 3 are determined by comparing the
ferromagnetic phase with the canted
ferromagnetic phase, the spin spiral phase,
the antiferromagnetic phase, and the phase separation in a mean field approach.

From the point of view of conduction electrons,
 the hopping of electrons is heavily
dragged by the spin background.
The hopping is prohibited when the angle $\Theta_{ij}=\pi$. The
effective interaction is also determined by $\Theta_{ij}$ as well as
$v_0/t$.
The ratio
$- (2v_0 \sin^2 \frac{\Theta_{ij}}{2})/(t \cos\frac{\Theta_{ij}}{2})$
approaches to zero when $\Theta=0$, i.e. the
ferromagnetic case, and becomes divergent when $\Theta=\pi$, i.e. the
antiferromagnetic case. The consequence is quite different from the
usual t-J model, in which the ratio $J/t$ is fixed and is usually very small.
When localized spins form a fully saturated ferromagnet,  the conduction
electrons are a spinless free fermion gas. Oppositely when the localized
spins form an antiferromagnetic background,  the attraction becomes rather
strong since the hopping of electrons are completely
suppressed even for very small $v_0/t$.
Strong attraction between fermions will lead to the instability to
the phase separation.
When the phase separation occurs, the system is divided into two parts:
electron-rich and -poor regimes.
In the electron-rich regime, all electrons
accumulate together and $\rho \rightarrow 1$. In this case, the kinetic
energy vanishes and the average energy per bond is
$-J_{AF}S^2 - 2 v_0$. In the electron-poor regime ($\rho=0$),
the average per bond is
$-J_{AF}S^2$. When $J_{AF}=0$, the spin background of the electron-poor
regime can be ferromagnetic. The phase separation arises in a very small regime near the half filling.
For a finite $J_{AF}$,  the spin background is antiferromagnetic. Hence the
average energy per bond for the whole system in the phase separation
is $\epsilon_{ps} = -(J_{AF}S^2 + 2 v_0 \rho)$, which is always lower than
the energy in an antiferromagnetic state with a uniform density of electrons.
This conclusion holds for any dimensional cases \cite{note-ps}.
Thus the phase separation occurs in the antiferromagnetic background.
The phase diagram in Fig. 1 shows that the phase separation occurs in the
case of either high or low density of electrons. Between the ferromagnetic
phase and the phase separation, it is a paramagnetic or non-collinear
magnetic regime.
The phase diagram is in good agreement with those established by utilizing
Monte-Carlo simulation by Yunoki {\it et  al.} \cite{Yunoki98}. The phase separation
always occurs near $\rho\rightarrow 1$ no matter how large $J_{AF}$ is
and $v_0 > 0$ because the antiferromagnetic coupling is
always predominant in the limit. This is quiet similar to those obtained in
the t-J model in the large S limit \cite{Auerbach91}.
Near $\rho \rightarrow 0$, $J_{AF}$ will determine whether the phase
separation
arises as the ferromagnetic coupling will dominates over the
antiferromagnetic coupling if $J_{AF}=0$. When $J_{AF}\neq 0$,
the phase separation can arise since the double exchange ferromagnetic
coupling is approximately proportional to the density of electron
near $\rho =0$, which is always less than a constant $J_{AF}$
at a sufficiently dilute doping. This is also consistent
with the numerical results \cite{Yunoki98}.
For a fixed $v_0$, the phase separation can occur when $J_{AF}$
increases as shown in Fig. 2.
If $v_0$ is very large,  the phase separation can occur for any $J_{AF}$,
and vice versa.If $v_0=0$, {\it i.e.}, the strong Hund coupling
the phase separation does not arise. Thus the attraction plays a decisive role
in the phase separation.
It is worth mentioning that  even in the paramagnetic phase the average value
of attraction potential $v_0 (\langle \cos \Theta_{ij} \rangle -1) = -v_0$,
half of that in the antiferromagnetic case. When the phase separation
occurs, it will
enhance the antiferromagnetic coupling especially in the electron-rich regime. This agrees with experimental observation
that the charge stripe arises at higher temperature than the
spin stripe by electron-diffraction and neutron-scattering measurements
\cite{Tranquada95}.

As a conclusion,  we discuss possible relevances of the attraction,
the phase separation, and the charge stripe phase. Although
the attraction could attract the electrons together to form an electron-rich
regime, other physical mechanism has to be taken into account in order to
explain the stripe behaviors for charge and spin. L\"{o}w {\it et  al}.
proposed the ideas of frustrated phase separation by considering the
nearest neighbor
attraction and long-range Coulomb repulsion \cite{Low94}.
Our work provides a direct
evidence that an attraction between electrons indeed arises from the
superexchange of electrons for the finite $J_H$ case. On the other hand,
the role of hopping term is still unclear in forming the stripe phase. Recent numerical
calculation by density-matrix renormalization group (DMRG) found some evidences
that the stripe behaviors appear in the two-dimensional t-J model
\cite{White98}
for a specific dopant due to the electron hopping even without the Coulomb
interaction. Our finding provides a direct mechanism for electrons
to condensate along the charge stripe. For example,  on a square lattice and at
$\rho=2/3$, a static charge stripe as shown in the inset of Fig.3 has a lower
energy than the state with a uniform density of charge for a larger
$v_0$. The larger $J_{AF}S^2$ is the lower the energy of the charge stripe is
for a fixed $v_0$. However the stripe state is unstable against the phase separation.
To stabilize the stripe state, one should
consider other physical processes, for example, effect of long-range Coulomb
interaction \cite{Low94} and non-collinear magnetism of localized spins. In short, since
the spin background modulates the electronic behaviors, 
we believe that our model (\ref{rde-model}) or its quantum form (\ref{tj-model})
is a good starting point to investigate the phase separation, the charge ordering
and spin ordering in doped transition-metal oxides. 

This work was supported by a CRCG research grant at the University of Hong
Kong.

\begin{figure}
%{\epsfxsize=\hsize \epsfbox{figure-1}}\caption
{The phase diagram for a square lattice at $J_{AF}S^2 = 0.05$. ``F''
means ferromagnetic. In that regime it is a metallic double exchange
ferromagnet. ``PS'' means the phase separation on an antiferromagnetic
background. ``NF'' is between the ferromagnetic phase and phase separation.
It is paramagnetic or mixture of some non-collinear
magnetism from the point of
view of the mean field theory.}
\label{fig-1}
\end{figure}

\begin{figure}
%{\epsfxsize=\hsize \epsfbox{figure-2}}
\caption{
The density dependence of the critical value of $J_{AF}S^2$ to the phase
separation for $v_0 = 0.05$, 0.10, and 0.02 in a square lattice.}
\label{fig-2}
\end{figure}

\begin{figure}
%{\epsfxsize=\hsize \epsfbox{figure-3} }
\caption{The $v_0$ dependence of the energy difference, E, between the charge
stripe and a state with a uniform density of charge for different
$J_{AF}S^2$. The inset is the charge stripe phase we discuss. The black
points stand for single occupancies of electrons.}        
\label{fig-3}
\end{figure}

\begin{figure}
{\epsfxsize=\hsize \epsfbox{figure-1}}
{\epsfxsize=\hsize \epsfbox{figure-2}}
{\epsfxsize=\hsize \epsfbox{figure-3}}
\end{figure}

%\end{multicols}


\begin{references}
\bibitem {note-ps}
It should be pointed out that the ratio of the densities in the
electron-poor
and -rich regimes are not always 0. We can say the phase separation occurs
as far as the ratio deviates from 1.
\end{references}
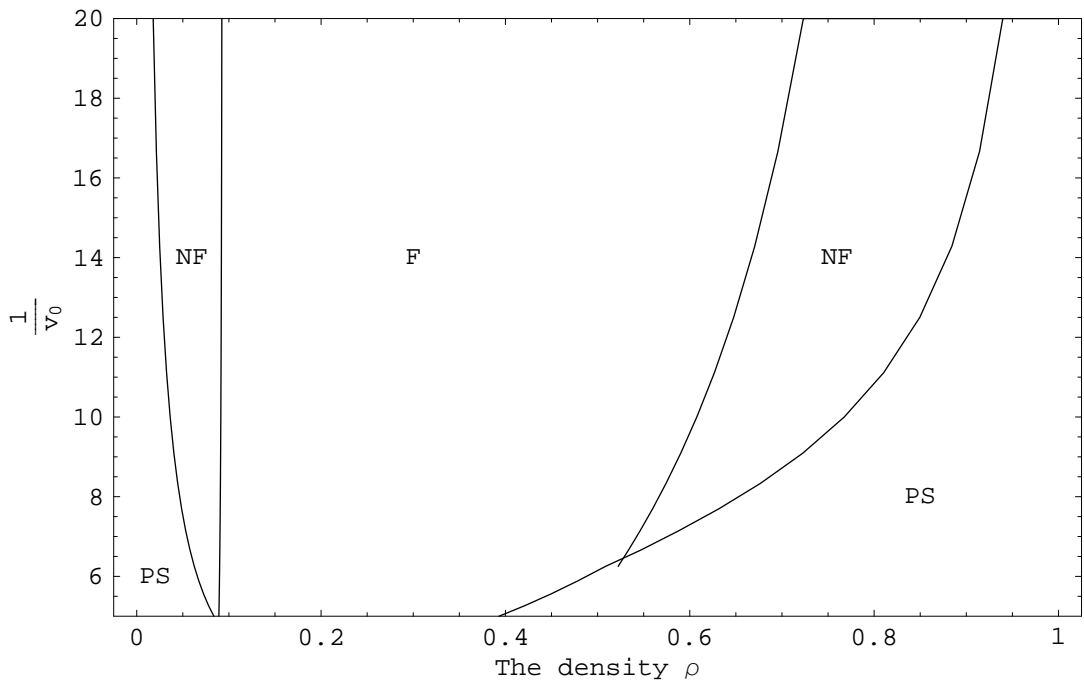
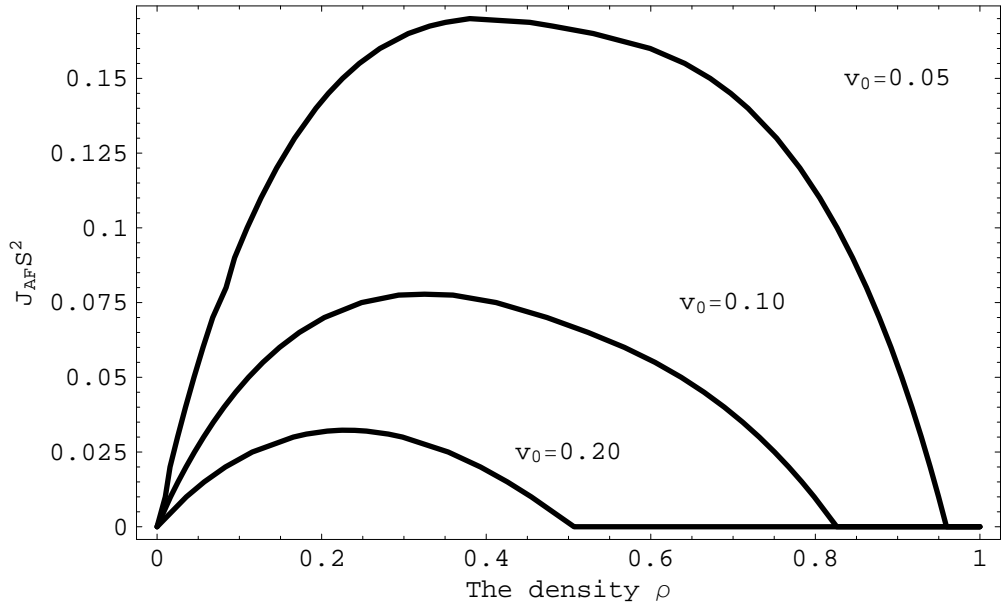
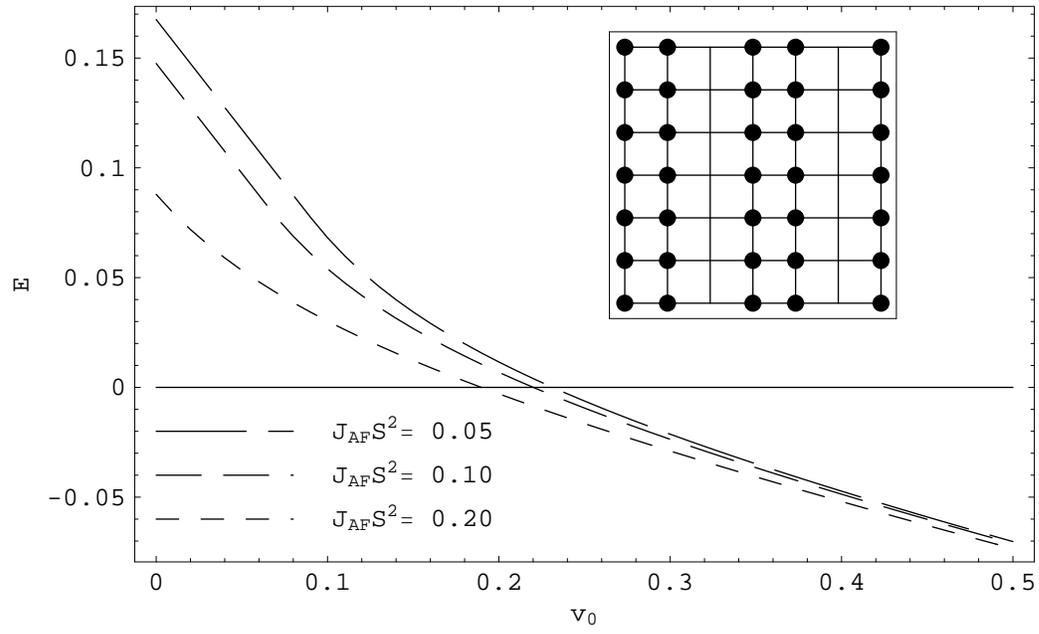
\end{document}